\documentclass[aps,reprint,superscriptaddress,nofootinbib]{revtex4-2}

\usepackage{graphicx}
\usepackage{dcolumn}
\usepackage{bm}
\usepackage{hyperref}
\usepackage[mathlines]{lineno}
\usepackage{amsmath} 

\usepackage{xcolor} 
\usepackage{footnote}
\usepackage{times}

\newcommand*\mean[1]{\overline{#1}}

\begin{document}

\title{
  Flattening a trapped atomic gas using a programmable optical potential in a feedback loop
}

\author{Sol Kim}
\affiliation{Department of Physics and Astronomy, Seoul National University, Seoul 08826, Korea}

\author{Kyuhwan Lee}
\affiliation{Department of Physics and Astronomy, Seoul National University, Seoul 08826, Korea}

\author{Jongmin Kim}
\affiliation{Department of Physics and Astronomy, Seoul National University, Seoul 08826, Korea}

\author{Y. Shin}
\email{yishin@snu.ac.kr}
\affiliation{Department of Physics and Astronomy, Seoul National University, Seoul 08826, Korea}
\affiliation{Institute of Applied Physics, Seoul National University, Seoul 08826, Korea}


\begin{abstract}
We present a method for producing a flat, large-area Fermi gas of $^6$Li with a uniform area density. 
The method uses a programmable optical potential within a feedback loop to flatten the \emph{in-plane} trapping potential for atoms. 
The optical potential is generated using a laser beam, whose intensity profile is adjusted by a spatial light modulator and optimized through measurements of the density distribution of the sample. 
The resulting planar sample exhibits a uniform area density within a region of about 480 $\mu$m in diameter and the standard deviation of the trap bottom potential is estimated to be $\approx k_{B} \times$ 6.1 nK, which is less than 20\% of the transverse confinement energy. 
We discuss a dimensional crossover toward 2D regime by reducing the number of atoms in the planar trap, 
including the effect of the spatial variation of the transverse trapping frequency in the large-area sample.
\end{abstract}

\maketitle

\section {Introduction}

In the realm of quantum phenomena within two-dimensional (2D) systems, the reduction in dimensionality engenders a diverse array of intriguing behaviors. 
Fractional quantum Hall states, characterized by the fractionalization of electron charges and the emergence of anyonic statistics, find roots in confined geometries~\cite{FQH-1, FQH-2}. 
Similarly, topologically ordered phases, typified by the toric code model, thrive in the planar expanse of 2D systems, allowing for the existence of nontrivial topological excitations \cite{Top-1}. 
Quantum criticality in 2D systems, manifesting even at absolute zero, results from enhanced quantum fluctuations inherent in lower dimensions~\cite{Qcriticality}. 
This reduced dimensionality provides a fascinating setting in which quantum coherence, entanglement, and novel phases flourish in the distinctive landscape of quantum many-body physics.

Extensive efforts have been made to explore the intricacies of 2D quantum phenomena in ultracold atom experiments \cite{2D-1, 2D-2, 2D-4, 2D-spin-imbalance-1, 2D-spin-imbalance-2, 2D-Fermi-2}. 
To address 2D physics, samples must be compressed unidirectionally, ensuring that the confinement energy exceeds relevant energy scales such as chemical potential $\mu$ and thermal energy. 
Compression is achieved by increasing one of the trapping frequencies in conventional 3D samples \cite{TEM01-1} or using one-dimensional optical lattices to create a stack of 2D samples~ \cite{lattice, accordion, 2DFermi2010}. 
Alternatively, one may envisage simply reducing the chemical potential, e.g., through evaporation comparable to or below the confinement energy in the transverse direction, instead of intensifying the transverse confinement.
However, in this case, it is necessary for the \emph{in-plane} trapping potential to be sufficiently flat over a substantial area for the sample to maintain its shape and uniform density profile even at low chemical potential in the 2D regime.
Efforts have been made to create a homogeneous 2D sample using a box potential~\cite{2D-Fermi-1}, which can be further improved by using a programmable one to compensate for any remaining inhomogeneity \cite{feedback-1d, feedback-2d}. 

In this paper, we demonstrate the production of a flat, large-area unitary Fermi gas of $^6$Li with disk geometry.
The trapping potential is homogenized \emph{in-plane} via feedback optimization of a programmable optical potential using a spatial light modulator (SLM) while remaining harmonic in the tightly confining transverse direction. 
The resulting flat sample retains its shape and density uniformity even with low atomic densities after deep evaporation. 
The standard deviation of the \emph{in-plane} trapping potential is estimated to be about \(k_{B} \times 6.1\)~nK over a region of 480 $\mu$m in diameter. The magnitude of the potential variations is less than 20\% of the transverse confinement energy, allowing us to reach a 2D regime by reducing the atomic density.

\begin{figure*}
  \includegraphics[width=7in]{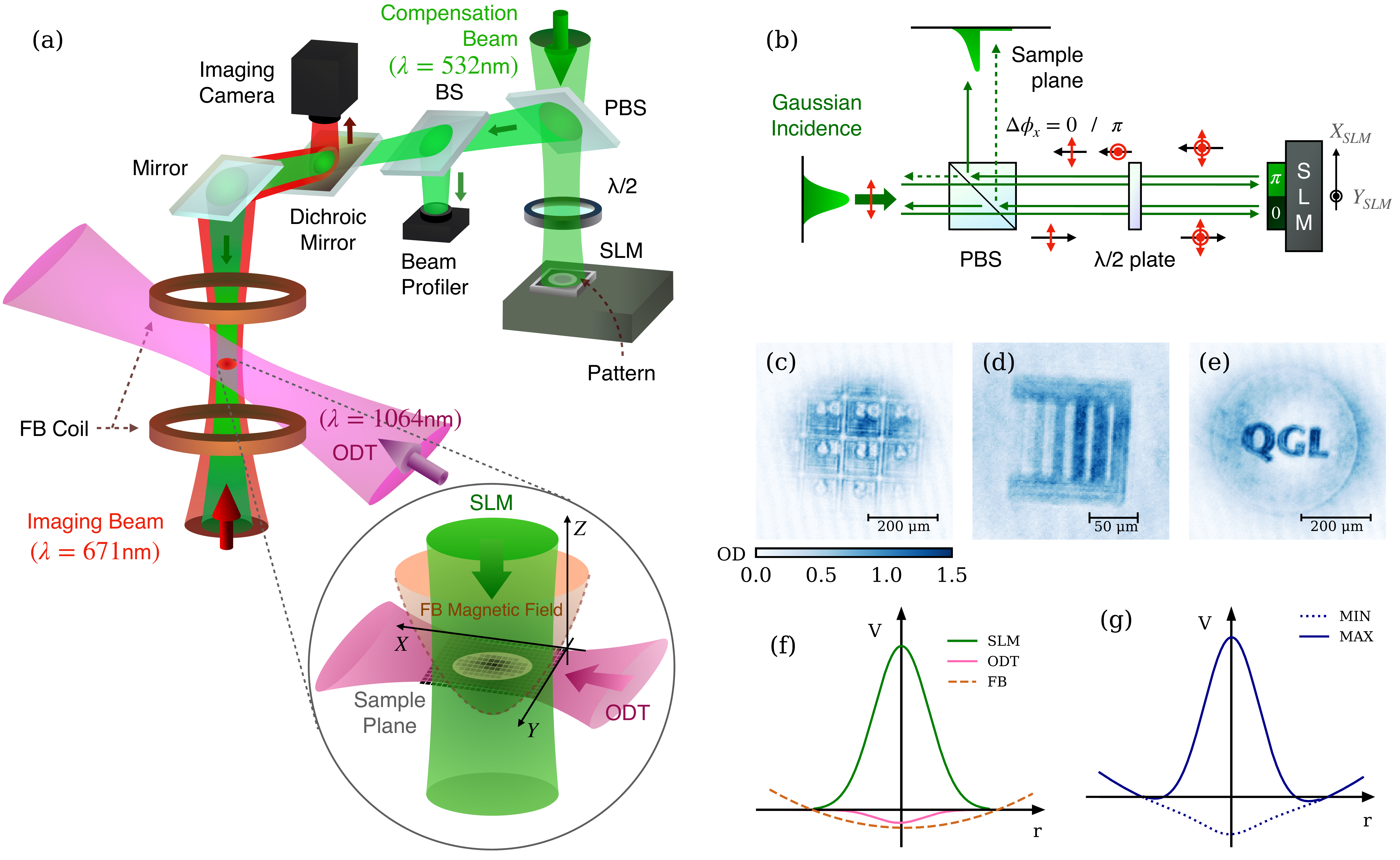}
  \caption{\label{fig:1} 
  Programmable trap for atoms.
(a) Schematic of the experimental setup. Atoms are confined in a hybrid trapping potential formed by an optical dipole trap (ODT), a Feshbach (FB) magnetic field, and a repulsive compensation beam (circular inset).
The compensation beam is irradiated along the axial direction of the sample and its intensity profile is controlled with a spatial light modulator (SLM).
(b) Illustration of the working principle of intensity modulation using the phase SLM. (c)-(e) Images of $^{6}$Li atom clouds trapped in various potentials, configured (c) in an alphanumeric grid pattern, (d) with multi-valued modulation, and (e) with our laboratory logo imprinted. 
(f) Radial distributions of the optical and magnetic potentials for the hybrid trap. (g) Net trapping potentials with the full compensation beam (solid line) and without the beam (dotted line).
  }
\end{figure*}

\section {Experiment}

\subsection{Sample preparation}

The schematic of our experimental apparatus is presented in Fig. 1(a).
As described in \cite{machine-thesis}, we prepare a degenerate Fermi gas of $^{6}$Li in a magnetic trap by sympathetic cooling with bosonic atoms of $^{23}$Na, and subsequently transfer it to an oblate optical dipole trap (ODT) formed by a 1064-nm laser beam.
The atomic sample is transformed into an equal mixture of the two lowest hyperfine spin states and further cooled via evaporation, where a magnetic field is tuned to the Feshbach resonance at 830 G for resonant interactions.

The evaporation is controlled using the ODT beam power, which determines the depth of the trap along the transverse direction to the trapping plane.
When the sample is evaporated to have a total number of atoms per spin state below approximately $1 \times 10^{6}$, it undergoes a superfluid transition, marked as a reference point for the transverse trap depth. 
Further cooling involves continued evaporation, resulting in a variable atom number depending on the final trap depth during the evaporation.
After this, the ODT beam power is increased to the reference value. 
In the final trapping condition, including the radial confinement due to the Feshbach field curvature, the radial trapping frequencies are $\{\omega_{x}, \omega_{y}\} \approx 2\pi \times \{18, 21\}~\text{Hz}$, while the transverse trapping frequency is $\omega_{z} \approx 2 \pi \times 700 \text{ Hz}$.

\subsection{Programmable trap}

To manipulate the \emph{in-plane} density distribution of the trapped sample, we apply an additional optical potential by irradiating a spatially tailored 532-nm laser beam along the transverse direction [Fig.~1(a)]. 
The laser beam generates a repulsive potential to $^6$Li and can compensate for the trapping of the ODT and the magnetic field [Fig.~1(f)].
The power of the SLM beam is approximately 1.2~W and the $1/e^2$ beam diameter is about 200 $\mu$m at the sample plane.
To modulate the intensity profile of the compensation laser beam in the sample plane, a liquid-crystal on silicon (LCoS) SLM (Meadowlark E-Series 1920×1200) is placed at the image plane of the sample for a 4-f imaging setup. 
The beam intensity is controlled using polarization optics as depicted in Fig.~1(b), where a polarizing beam splitter (PBS) and a half-wave plate are configured to rotate the linear polarization of the incident laser beam to a 45$^\circ$ inclination with respect to the SLM fast axis ($X_\text{SLM}$-axis), and the birefringent phase shift $\Delta\varphi_{x}$ by the SLM determines the intensity of the reflected beam from the PBS. 
This results in an overall transmission ratio of $\sin^{2}\left(\Delta\varphi_{x}/2\right)$, which can be programmed in pixel-wise grayscale over the SLM plane.

In Figs.~1(c)-(e), we show various in-plane density distributions of trapped atomic samples.
Compared with a conventional holographic setup using the SLM~\cite{DMD-fourier, SLM-fourier-1, SLM-fourier-2}, this setup does not produce holographic speckles; therefore, it has a relatively large working area with high quality. 
Furthermore, compared to a different approach using a digital micromirror device (DMD), which is operated in a switching mode, this SLM method allows for inherent grayscale control without loss of resolution~\cite{DMD-err-diff,affine-1} and heating~\cite{DMD-pwm-mtf}.

\begin{figure}
  \includegraphics[width=3.4in]{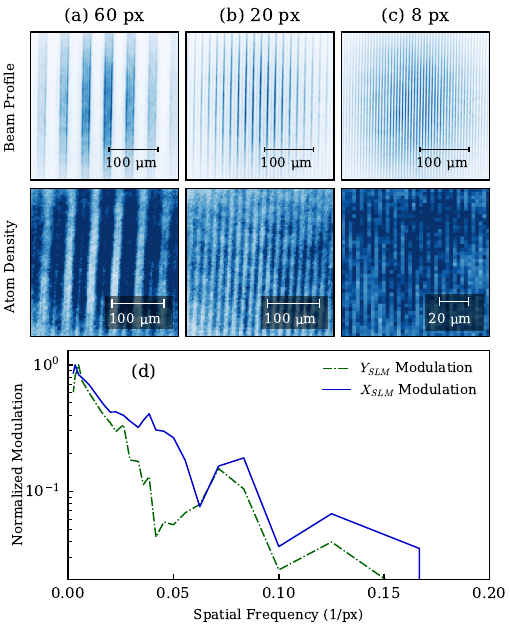}
  \caption{\label{fig:2} Characterization of the spatial resolution of the programmable trap.
The compensation beam is square-modulated with different periods of (a) 60, (b) 20, (c) 8 pixels on the SLM plane. The upper and lower rows show the corresponding images of the beam intensity, obtained by the beam profiler (see Fig.~1), and those of the resulting atomic density.
(d) Normalized amplitude of density modulations as a function of the spatial frequency of the intensity modulations of the compensation beam.}
\end{figure}

The SLM laser beam is irradiated on the sample using the imaging system which is also used for the absorption imaging of the sample [Fig.~1(a)]. Due to the limited imaging resolution, the actual optical potential delivered onto the sample is spatially smoothed because the imaging system filters out high-frequency components. 
To evaluate the resolution of the imaging system, we illuminate a one-dimensional square wave potential with various wavelengths using the SLM (see Fig.~2 first row) and capture an absorption image of the sample to measure the resulting modulations in the density distribution (see Fig.~2 second row)~\cite{DMD-pwm-mtf}. 
The modulation transfer efficiency is determined by dividing the image by its blurred counterpart for normalization and then calculating the standard deviation of the relative optical signal. The results are shown in Fig. 2(d). 
The relative modulation intensity decreases as the spatial frequency of the periodic potential increases, and falls significantly below $5\%$
when the spatial frequency exceeds 0.125 (SLM pixel)$^{- 1}$, corresponding to 6~$\mu$m in the sample plane.
This threshold value is considered as the resolution limit of our imaging system and therefore the controllable limit for our programmable trap.
For the analysis of density distribution, we apply low-pass filtering to images with this cutoff frequency (see Appendix B).

\begin{figure}
  \includegraphics[width=3.4in]{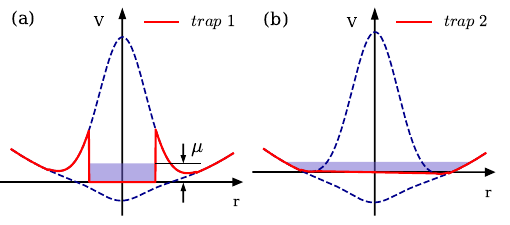}
  \caption{\label{fig:3} 
  Homogenization strategy. The trap bottom potential is flattened 
(a) with a ring-shaped wall potential (\textit{trap 1}) and
(b) without the wall (\textit{trap 2}). 
The dashed lines indicate the dynamic range of the programmable trapping potential [Fig. 1(g)]. The blue-shaded region indicates the chemical potential $\mu$ of a trapped sample.
  }
\end{figure}

\subsection{Flat sample generation}

In this work, we demonstrate the preparation of planar samples with uniform area densities using the programmable trap. This is achieved by balancing the radial attraction from the ODT and the Feshbach field curvature with the repulsive potential of the compensation laser beam from the SLM.
Two different trap configurations are used, as illustrated in Figs.~3(a) and 3(b); one with a ring-shaped boundary (referred to as \emph{trap 1}) and the other without it (referred to as \emph{trap 2}). The wall potential is formed by maximizing the intensity of the SLM beam in its outer part [Fig.~3(a)]. In the case of \emph{trap 2} without a wall potential, the area of the sample can be enlarged to the limit allowed by the geometry and power of the SLM laser beam, which compensates for the radial trapping force of the Feshbach field curvature.

The spatial profile of the SLM beam is optimized using a feedback technique based on the measured \emph{in-situ} column density distribution $n_{\text{2D}}(x,y)$ of a trapped sample.
A feedback loop is initiated by taking an absorption image of the \emph{in-situ} density distribution and preprocessing it to eliminate noise and defects. Subsequently, FFT filtering is applied to remove interference patterns caused by vibrations of the experimental setup, and then a low-pass filter is used to decrease shot noise (see Appendix B). 
By comparing the processed image with the desired density distribution, a new phase profile for the SLM is calculated and then transmitted to the modulator plane through an affine transformation~\cite{affine-2}. For the calculation of phase adjustments based on the error signal, a fuzzy logic feedback system is adopted~~\cite{fuzzy-1}, whose details are provided in Appendix C.
Through multiple iterations of this feedback process, the absorption image gradually aligns with the target density distribution [Figs. 4(a)-(d)].

\subsection{Atom column density in the planar trap}

We characterize our planar trap by modeling its potential as 
\begin{equation}
    V(x,y,z) = V_\perp(x,y) + \frac{1}{2} m \omega_z (x,y)^2 z^2,
\end{equation}
where \( V_\perp (x,y)\) represents the trap bottom potential along the sample plane and $m$ is the atomic mass. Without loss of generality, we assume that the mean value of the trap bottom potential is $\mean{V_\perp}=0$.
We note that the transverse confinement may vary spatially as $\omega_z(x,y)$, which is more likely when the area of the sample is large.

Here we describe the relation of the column density $n_\text{2D}$ to the trap parameters $V_\perp$ and $\omega_z$ at zero temperature, which will be used in our subsequent analysis of the experimental data.
In the 3D regime for $\mu \gg \hbar \omega_z$, using the Thomas-Fermi approximation,
the local atom density per spin state is given by $n_{\text{3D}} = \frac{1}{6\pi^2} (2m \mu_\text{local} / \xi \hbar^2)^{3/2}$, where $\mu_\text{local} = \mu - V$ is the local chemical potential and $\xi$ is the Bertsch parameter for the unitary Fermi gas~\cite{zwierlein-book}. 
Integrating the density along the transverse direction yields the 2D column density per spin state as 
\begin{equation}
n_{\text{2D}}(x,y) = \frac{m}{4\pi \hbar^3\xi^{3/2}} \frac{[\mu-V_\perp (x,y)]^2}{ \omega_z(x,y)}.    
\end{equation}
It is clear that $n_\text{2D}$ is affected by both trap parameters $V_\perp$ and $\omega_z$.
When the variations of $V_\perp$ and $\omega_z$ in the sample plane are not significant, the chemical potential is approximated as 
\begin{equation}
    \mu = \sqrt{\frac{4\pi \hbar^3 \xi^{3/2} }{m} \mean{\omega_z}~\mean{n_\text{2D}} }.
\end{equation}
where $\mean{n_\text{2D}}$ and $\mean{\omega_z}$ stand for the mean values of $n_{\text{2D}}$ and $\omega_z$, respectively.

As the chemical potential decreases to $\mu \sim \hbar \omega_z$, the relation of $n_\text{2D}$ and $\{V_\perp,\omega_z\}$ is changed from Eq.~(2). In the ideal 2D regime, where the $z$ directional motion of the atom is completely restricted to the ground state of the transverse harmonic potential, the Fermi momentum in the trap plane is given by $k_{F} = \sqrt{4\pi n_{\text{2D}}}$. Then, the local chemical potential is $\mu_\text{local} = \xi(\hbar^2 k_F^2/2m)+\hbar \omega_z / 2=
\mu- V_\perp$, including the zero-point energy due to the transverse confinement, and the area density of atoms in the 2D system is given by 
\begin{equation}
    n_{\text{2D}}(x,y) =\frac{m}{2 \pi \hbar^2 \xi} \left[ \mu -V_\perp -\frac{1}{2}\hbar \omega_z\right]. 
\end{equation}
In the 2D regime, the chemical potential is expressed as
\begin{equation}
    \mu = \frac{2\pi \hbar^2 \xi }{m} \mean{n_{\text{2D}}} + \frac{1}{2}\hbar \mean{\omega_z}.
\end{equation}

\begin{figure}
  \includegraphics[width=3.4in]{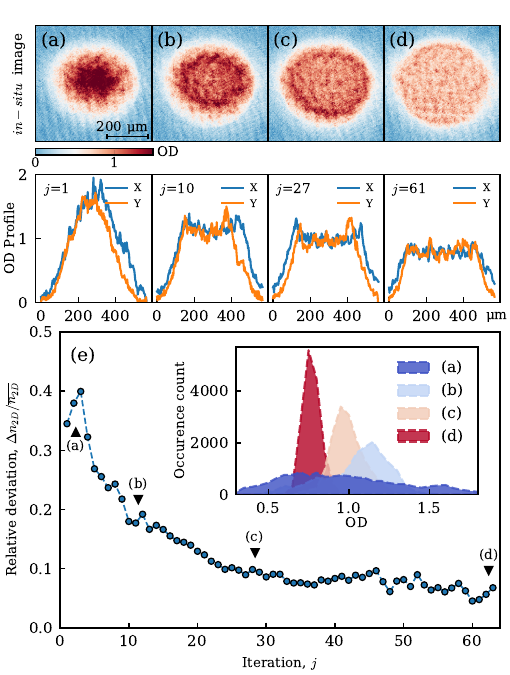}
  \caption{\label{fig:4} 
  Feedback homogenization of atomic Fermi gas.
  (a)-(d) \emph{in-situ} density images for different numbers of feedback iterations, $j$, and the corresponding density profiles across the center along the horizontal (X) and vertical (Y) directions. \emph{Trap 1} is used [Fig.~3(a)].
  (e) Relative density deviation $\Delta n_\text{2D}/\mean{n_\text{2D}}$ versus feedback iteration number $j$, where $\mean{n_\text{2D}}$ and $\Delta n_\text{2D}$ are the mean and the standard deviation of the atom density $n_\text{2D}$ inside the ring wall. 
  The inset shows the occurrence plots of $n_\text{2D}$ in (a)-(d).}  
\end{figure}

\section{Results}

\subsection{Feedback homogenization of the atomic column density}

Figures 4(a)-(d) show the evolution of the density distribution of the sample in the feedback homogenization process using \emph{trap 1}.
Through the iterative application of feedback optimization, it is evident that the \emph{in-situ} density profile flattens. 
In particular, the central region, with a wider dynamic range of the SLM beam intensity, exhibits faster convergence, leading to an expansion of the uniformed region with each interaction.
In Fig.~4(e), The evolution of the relative density deviation $\Delta n_\text{2D}/\mean{n_\text{2D}}$ is presented, where $\Delta n_\text{2D}$ is the standard deviation of the column density $n_\text{2D}$ in the flat central zone that is a circular region of 360~$\mu$m in diameter. It is observed that the relative density deviation is decreased to below 10\% after approximately 30 feedback loops. 
Subsequent iterations lead to further enhancement of density uniformity, as shown in Fig. 4(d).
The inset of Fig.~4(e) shows the occurrence distribution of optical density (OD), which becomes narrower and more pronounced with increasing $j$ (number of iterations), indicating the flattening of the trap bottom.
Generally, in the experiment, achieving a homogeneity quality level comparable to that in Fig.~4(e) requires around 50 iterations.

After completion of feedback homogenization, the relative column density deviation is reduced to $3.98\%$ with $\mean{n_{2D}} = 3.40~\rm{\mu m}^{-2}$.
For \emph{trap 2}, we obtain $\Delta n_\text{2D}/\mean{n_\text{2D}}=8.22\%$ with $\mean{n_{2D}} = 2.60 ~\rm{\mu m}^{-2}$, where the flat region is elliptical with a major axis of 488 $\mu$m and a minor axis of 408 $\mu$m (Fig.~6).
Relative density fluctuations are higher in \emph{trap 2} than in \emph{trap 1}, and this degradation in feedback performance is due in part to lower $\mean{n_\text{2D}}$. It is also observed that a radial ring wall at the sample boundary facilitates the redistribution of atoms within a specific region, thereby improving the quality of feedback.

\subsection{Estimation of the trap bottom flatness}

According to the relation of Eq.~(2), the variation of the column density is described as
\begin{equation}  \label{reldev-3D}
    \frac{\delta n_{\text{2D}} (x,y)} { \mean{n_{\text{2D}}} } = - 2 \frac{\delta V_\perp (x,y) }{\mu} - \frac{\delta \omega_z(x,y)} {\mean{\omega_z}}.
\end{equation}
Because $V_\perp$ has been spatially adjusted in feedback homogenization to minimize $\delta n_\text{2D}$, $\delta V_\perp$ can be decomposed as
\begin{equation}  
    \delta V_\perp= \delta V_{\perp,z} + \delta V_{\perp,0},
\end{equation}
where $\delta V_{\perp,z}=-\frac{\mu_\text{ho}}{2 \mean{\omega_z}}\delta \omega_z$ reflects the effect of $\delta \omega_z$ and $\delta V_{\perp,0}=-\frac{\mu_\text{ho}}{2}\frac{\delta n_\text{2D}}{\mean{n_\text{2D}}}$ represents the residual part resulting from the technical limit of feedback optimization. Here the subscript `ho' denotes the value for the homogenized sample. 
It is reasonable to assume that $\delta V_{\perp,0}$ is uncorrelated with $\delta \omega_z$, and the flatness of the trap bottom potential can be estimated with its standard deviation as
\begin{equation}
    \Delta V_\perp=\sqrt{\Delta V_{\perp, z}^2 + \Delta V_{\perp, 0}^2}= \frac{\mu_\text{ho}}{2} \sqrt{\sigma_z^2+ \sigma_n^2},
\end{equation}
where $\sigma_z=\Delta \omega_z/\mean{\omega_z}$ and $\sigma_n=(\Delta n_\text{2D}/\mean{n_\text{2D}})_\text{ho}$.

The spatial distribution of the transverse trapping frequency $\omega_z(x,y)$ is investigated using a parametric heating method, in which the intensity of the ODT beam is periodically modulated with frequency $\omega_m$ and the parametric resonance would occur at $\omega_m=2\omega_z$, resulting in loss of atoms due to heating~\cite{parametric}.
In particular, to address the spatial inhomogeneity of $\omega_z$, we additionally apply a grid-patterned wall potential created by the programmable SLM laser beam [Fig.~5(a) inset]. The height of the grid wall potential is significantly higher than the chemical potential, so that individual cells are isolated from their surroundings. This enables us to measure local atom loss and determine the resonant frequency for each cell of the grid [Fig. 5(a)].

In Figs.~5(b) and 5(c), we show the measurement results of $\omega_{z}(x,y)$ before and after homogenization, respectively. 
The spatial distribution of $\omega_z$ reveals a dipole-like structure across the sample, which we ascribe to imperfections in the ODT.
The relative deviation is measured as $\sigma_z=\Delta \omega_z/\mean{\omega_z}= 2.09\%$.
In the homogenized trap, the mean value of $\omega_z$ is observed to increase by $2\pi \times$ 40~Hz from that in the bare trap. 
This change could be attributed to the divergence of the SLM laser beam as it traverses the trapping plane of the ODT.

From Eq.~(8) and with the measured values of $\sigma_n$ and $\sigma_z$, we obtain $\Delta V_\perp \approx k_\text{B} \times 3.5$~nK for \emph{trap 1} and $\approx k_\text{B} \times 6.1$~nK for \emph{trap 2}. Here, the values of $\mu_{\text{ho}} = k_\text{B}\times 163\text{ nK}$ (\emph{trap 1}) and $k_\text{B}\times 143\text{ nK}$ (\emph{trap 2}) are calculated using Eq.~(3) based on the measured values of $\mean{n_\text{2D}}$ and $\mean{\omega_z}$.
$\Delta V_\perp$ within the homogenized trap is considerably smaller than the transverse confinement energy of $\hbar\mean{\omega_z}=k_\text{B}\times 34$~nK, 
indicating that our system could maintain its homogeneity when entering a 2D regime for $\mu\sim \hbar \mean{\omega_z}$.

\begin{figure}
  \includegraphics[width=3.4in]{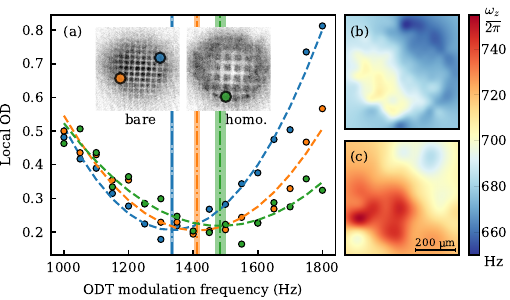}
  \caption{\label{fig:5} 
  Spatial variation of the transverse trapping frequency $\omega_z$.
  (a) Local atom loss is measured with a grid-wall potential as a function of the ODT modulation frequency (see the text for details). 
  The left (right) image in the inset shows the atom density distribution under the grid-wall potential without (with) feedback homogenization.  
  The colored circles indicate the positions of the cells in the grid, corresponding to the three measurement data sets shown in (a).
  The dashed line indicates a quadratic function fit to a data set and the vertical dashed-dot line indicates the resonance frequency. The shade region denotes $1\sigma$ fit uncertainty.
  (b) Spatial distribution of $\omega_{z}$ for the bare trap before homogenization and (c) that for the homogenized trap.
}
\end{figure}

\begin{figure}
  \includegraphics[width=3.4in]{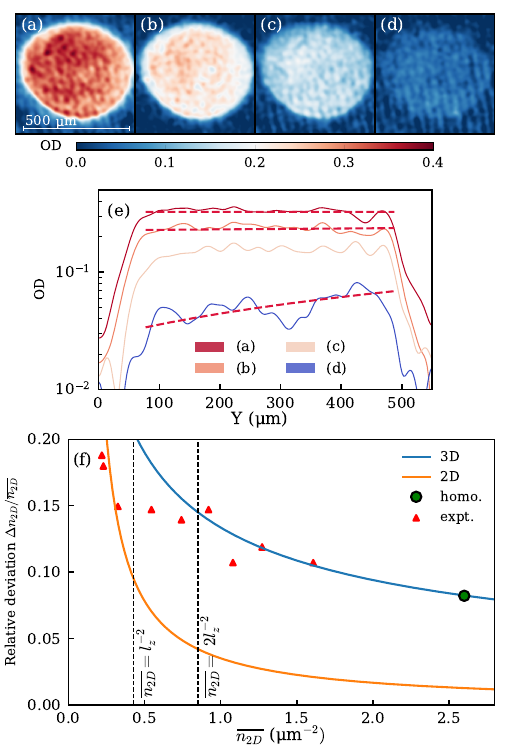}
  \caption{\label{fig:6} 
  Planar Fermi gas in the 3D-to-2D crossover.
  (a)-(d) {\it in-situ} images of $^6$Li clouds for various atom numbers and (e) the corresponding density profiles across the center along the vertical (Y) direction. The dashed lines indicates linear fits to the data in the center region.
  (f) Relative density deviation $\Delta n_\text{2D}/\mean{n_\text{2D}}$ for different mean area density $n_\text{2D}$.
  The blue and yellow lines indicate the 3D and 2D model curves of Eqs.~(10) and (11), respectively, for the measured values of the trap potential parameters. 
  The green circle denotes the sample condition at the feedback homogenization.
  $l_z = \sqrt{\hbar/m \mean{\omega_z}}$ is the harmonic oscillator length for the transverse confining. In the dilute regime for $\mean{n_\text{2D}}<2l_z^{-2}$, the experimental data deviate from the 3D model, indicating a crossover to 2D. 
  }
\end{figure}

\subsection{Toward 2D regime}

Using homogenized \emph{trap 2}, we explore the crossover to the 2D regime by reducing the number of atoms of a trapped sample \cite{3d-2d-crossover}.
As previously mentioned, the atom number is regulated by adjusting the lowest depth of the ODT during evaporation, and after the evaporation cooling, the ODT power is adjusted back to the level used during the feedback homogenization procedure. 
In Figs.~6(a)-(d), some of the resulting images are presented, along with the density profiles cut in their centers in the $Y$ direction shown in Fig. 6(e). 
These images demonstrate that the sample maintains a flat density profile even after substantial evaporation. For the lowest atom number, the column density is estimated to be $\mean{n_\text{2D}}= 0.22~\mu\text{m}^{-2}$, which is below the characteristic area density $n_{\text{2D},c}=l_z^{-2}\approx 0.4~\mu\text{m}^{-2}$ for the 2D regime. Here, $l_z =\sqrt{\hbar/m\mean{\omega_{z}}} \approx 1.5~\mu$m represents the harmonic oscillator length for transverse trapping. 
Additionally, it is observed that the density profile develops a slight slope as the atom density decreases, and it is noted that the direction of its gradient aligns with the observed variation of $\omega_z$ across the sample as per Eq.~(6).

In Fig.~6(f), the experimental results of $\Delta n_{\text{2D}} / \mean{n_{\text{2D}}}$ are presented for various atom densities.
It is observed that as the column density decreases from 2.60~$\mu\text{m}^{-2}$ to 0.22~$\mu\text{m}^{-2}$, the magnitude of relative density fluctuations increases gradually from 8\% to 19\%.
This behavior of the relative density deviation is understandable from the dependence of $\delta n_\text{2D}/\mean{n_\text{2D}}$ on $\mu$ in Eq.~(6). Note that in this work we exclude the shot noise effect because of the low-pass filtering applied during image processing.

For a quantitative understanding of the experimental data, we analyze how $\delta n_\text{2D}$ changes as $\mu$ varies, while keeping $\delta V_{\perp,0}$ and $\delta \omega_z$ constant.
For $\delta V_{\perp,z}=-\frac{\mu_\text{ho}}{2 \mean{\omega_z}}\delta \omega_z$ and $\delta V_{\perp,0}=-\frac{\mu_\text{ho}}{2} (\frac{\delta n_\text{2D}}{\mean{n_\text{2D}}})_\text{ho}$, Eq.~(6) is rewritten as 
\begin{equation} 
    \frac{\delta n_{\text{2D}}} { \mean{n_{\text{2D}}} } = \frac{\mu_\text{ho}}{\mu} \left(\frac{\delta n_\text{2D}}{\mean{n_\text{2D}}}\right)_\text{ho} + \left( \frac{\mu_\text{ho}}{\mu} -1 \right) \frac{\delta \omega_z}{\mean{\omega_z}},
\end{equation}
and the relative density deviation is given by 
\begin{equation}
    \frac{\Delta n_\text{2D}}{\mean{n_\text{2D}}} = \frac{1}{\sqrt{\nu}}\sqrt{ \sigma_n^2 + (1-\sqrt{\nu})^2 \sigma_z^2} 
\end{equation}
with $\nu=\mean{n_\text{2D}}/(\mean{n_\text{2D}})_\text{ho}$. 
Similarly, in 2D, Eq.~(4) gives
\begin{eqnarray} \label{reldev-2D}
    \delta n_{\text{2D}} &=& 
    -\frac{m}{2\pi \hbar^2 \xi} \left( \delta V_\perp + \frac{1}{2}\hbar\delta\omega_{z} \right) \nonumber \\
    &=& \frac{m \mu_\text{ho}}{4\pi \hbar^2 \xi} 
    \left[ \left(\frac{\delta n_\text{2D}}{\mean{n_\text{2D}}}\right)_\text{ho} + \left(1-\frac{\hbar\mean{\omega_z}}{\mu_\text{ho}}\right) \frac{\delta \omega_z}{\mean{\omega_z}}\right],
\end{eqnarray}
resulting in 
\begin{equation} 
    \frac{\Delta n_{\text{2D}}} { \mean{n_{\text{2D}}} }  = \frac{\sqrt{\xi \nu_c} }{\nu}  
    \sqrt{
    \sigma_n^2 + (1-\sqrt{\nu_c})^2 \sigma_z^2
    },
\end{equation}
where $\nu_c=n_{\text{2D},c}/(\mean{n_\text{2D}})_\text{ho}=(\hbar\mean{\omega_z}/\mu_\text{ho})^2$.

In Fig.~6(f), we provide the 3D and 2D model curves of Eqs.~(10) and (12), respectively, with $\sigma_n =8.22\%$, $\sigma_z =2.09\%$, $\nu_c=0.0565$ and $\xi = 0.37$, along with the experimental data. 
Interestingly, the measured values of $\Delta n_{\text{2D}} / \mean{ n_{\text{2D}}}$ are consistent with the 3D predictions for $\mean{n_{\text{2D}}} > 2/l_{z}^{2}$, and with the 2D predictions for $\mean{n_{\text{2D}}} < 1/l_{z}^{2}$. 
This observation indicates the crossover of the Fermi gas system from 3D to 2D as $\mean{n_\text{2D}}$ decreases. 
It is important to note that the Bertsch parameter may vary in the dimensional crossover. A recent experimental study suggested a value of $\xi\approx 0.2$ in a strongly 2D regime~\cite{2deos}.

\section{Summary}

We have introduced a programmable trap for atoms, using a SLM positioned at the image plane of the atoms, and demonstrated a feedback method to homogenize the column density of a planar sample over a substantial area of about 480 $\mu$m in diameter. 
Through an analysis of density deviations for various sample conditions, we have estimated that the roughness of the trap bottom potential is less than 20$\%$ of the transverse confinement energy and consequently observed that the sample maintains the area density homogeneity even as it approaches the 2D regime with low number of atoms.

By further refining the uniformity of transverse confinement and optimizing the feedback mechanism, we expect that this programmable trap could provide an ideal setting suitable for a quantitative study of the thermodynamics and dynamic properties of strongly interacting Fermi gases in the large space of system parameters including interaction strength and population imbalance between the spin components~\cite{zwierlein-book}, as well as spatial dimensionality~\cite{2deos,2deos-vale,3d-2d-crossover,quantum-anomaly-2,quantum-anomaly}. Recently, using feedback-homogenized planar samples, we have investigated the Kibble-Zurek mechanism for the superfluid phase transition~\cite{Lee24}.

\begin{acknowledgments}
We thank Taehoon Kim for experimental assistance. This work is supported by the National Research Foundation of Korea (Grants No. NRF-2023M3K5A1094811 and No. NRF-2023R1A2C3006565).
\end{acknowledgments}

\appendix 

\section {Position mapping between the SLM and the camera}

The precise position mapping between the SLM and the imaging camera is crucial for the reliable performance of the programmable trap, as even a slight misalignment can adversely affect the feedback convergence.
To establish the precise position mapping, we use an alphanumeric grid pattern displayed on the SLM, consisting of unique alphanumeric identifiers for each grid point. By correlating these identifiers with the corresponding grid points in the atom absorption image [Fig.~1(c)], we determine the mapping between individual pixels on the SLM and the camera.
This mapping is derived using three grid points that form a largest right-equilateral triangle to calculate the affine transformation, which considers factors such as magnification, rotation, reflection, and translation, while assuming a flat imaging plane without considering curvature effects~\cite{affine-1}. Our imaging setup allows us to pinpoint the position of a grid point in the absorption image with an accuracy of within 2 pixels, and any errors in the mapping are insignificant compared to the scale of our low-pass filtering (Appendix B.2).
In the imaging setup, the magnification from the modulator to the sample plane is 32:3, and the absorption image of a sample is taken at a magnification of 2.5.
The pixel sizes of the SLM device and the imaging CCD are 8~$\mu$m and 6.45~$\mu$m, respectively.

\section {Image preprocessing}

During the feedback homogenization process, the absorption image of the sample is preprocessed before being used to produce the feedback output. Initially, the background interference fringe pattern caused by machine vibrations is eliminated through FFT filtering, which involved filtering out the Fourier components of the image corresponding to the specified wave vector of the fringes. However, for the analysis depicted in Figs. 4 and 6, the FFT filtering is intentionally omitted to ensure a conservative estimation of $\Delta n_\text{2D}/\mean{n_\text{2D}}$. 
The influence of the fringe pattern on the measured values of $\Delta n_\text{2D}$ is minimal, accounting for less than a few percent of $\mean{n_\text{2D}}$.

Subsequently, a low-pass filter is applied on the absorption image, taking into account the estimated imaging resolution of approximately $6~\mu$m, equivalent to a length of 2.33 pixels on the camera plane. The images are Gaussian-blurred three times with a standard deviation of 2 pixels. With an effective averaging area of about 18 pixels, the shot noise would be reduced by a factor of $\sqrt{18}\approx 4.2$ due to the binning effect. Given the clarity of its effects and rationale, low-pass filtering is commonly employed for both homogenizing iterations and primary data analysis.

\section {Details of feedback homogenization}

In our feedback homogenization process, a fuzzy logic feedback approach is adopted, which is recognized for its resilience in scenarios where the error signal is affected by noise and susceptible to fluctuations~\cite{fuzzy-1}. This approach involves producing a feedback output in steps, taking into account the magnitude of the error signal, and disregarding minor fluctuations that could potentially cause feedback overrun. The criteria to produce feedback output in our operation are fine-tuned based on empirical data to ensure optimal convergence, as detailed in Table 1.

\begin{table}[h]
  \begin{center}
      \begin{tabular}{ |c|c| } 
       \hline
       Relative OD difference, \% & SLM phase jump, bit\\ 
       \hline
       $>$ 50\%   & 10  \\ 
       \hline
       $>$ 30\%   & 5 \\ 
       \hline
       $>$ 20\%   & 3  \\ 
       \hline
       $>$ 10\%   & 1  \\ 
       \hline
       $<$ 10\%   & 0  \\ 
       \hline
      \end{tabular}
      \caption {\label{table1} Fuzzy logic table for feedback homogenization.
      As an error signal, we calculate the relative OD difference from the absorption image, compared to the target OD profile for each pixel position. Note that a single bit of phase jump corresponds to the birefringent phase shift of approximately $\pi/128$.}
  \end{center}
\end{table}

In some cases, the trap bottom potential may gradually increase over feedback iterations while maintaining its uniformity. This could lead to atoms spilling over the boundary wall in \emph{trap 1} or to a decrease in the sample area in \emph{trap 2}. This is prevented by manual intervention in the feedback process, if necessary, where the target OD value is slightly adjusted by scaling it with a constant factor ranging from 0.9 to 1.1.

The phase profile for the SLM is low-pass filtered before being sent to the SLM. Since the error signal used for the feedback was low-pass filtered, there is a possibility that high-frequency elements might persist in the SLM phase profile, causing the feedback process to potentially converge towards inaccurate solutions. Aligning the low-pass filtering of the SLM phase profile with that of the error signal, in terms of the cutoff frequency, improved the quality of the feedback.

\end{document}